\documentclass[3p,twocolumn]{elsarticle}
\usepackage{siunitx}

\usepackage{lineno}
\modulolinenumbers[5]

\usepackage{color}

\journal{Fusion Engineering and Design, part of SOFT 2014}

\begin{document}

\begin{frontmatter}

  \title{Design and development activities for in-vessel and in-port
    components of ITER microwave diagnostics}

\author[ITER,Fircroft]{Antoine~Sirinelli\corref{mycorrespondingauthor}}
\cortext[mycorrespondingauthor]{Corresponding author}
\ead{Antoine.Sirinelli@iter.org}

\author[RFDA]{Nikolay~Antonov}
\author[PPPL]{Russel~Feder}
\author[ITER]{Thibaud~Giacomin}
\author[ORNL]{Gregory~Hanson}
\author[PPPL]{David~Johnson}
\author[NRC]{Vitaliy~Lukyanov}
\author[Bertin]{Philippe~Maquet}
\author[ITER]{Alex~Martin}
\author[TuV]{Johan~W.~Oosterbeek}
\author[ITER]{Christophe~Penot}
\author[ITER]{Micka\"el~Portal\`es}
\author[PPPL]{Catalin~Roman}
\author[F4E]{Paco~Sanchez}
\author[NRC]{Dmitry~Shelukhin}
\author[ITER]{Victor~S.~Udintsev}
\author[ITER]{George~Vayakis}
\author[NRC]{Vladimir~Vershkov}
\author[ITER]{Michael~J.~Walsh}
\author[PPPL]{Ali~Zolfaghari}
\author[RFDA]{Alexander~Zvonkov}

\address[ITER]{ITER Organization, Route de Vinon sur Verdon, CS 90 046, 13067 St. Paul Lez Durance Cedex, France}
\address[Fircroft]{Fircroft, Lingley House, 120 Birchwood Point, Birchwood Boulevard, Warrington, WA3 7QH, UK}
\address[RFDA]{Project Center ITER, Moscow, Russian Federation}
\address[PPPL]{Princeton Plasma Physics Laboratory, Princeton, NJ, USA}
\address[ORNL]{Oak Ridge National Laboratory, Oak Ridge, TN, US}
\address[NRC]{National Research Centre "Kurchatov Institute'', Moscow, Russia}
\address[Bertin]{Bertin Technologies, P\^ole d'activit\'es d'Aix-en-Provence, 155 rue Louis Armand, CS 30495, 13593 Aix-en-Provence, France}
\address[TuV]{Eindhoven University of Technology, Den Dolech 2, 5612 AZ Eindhoven, Netherlands}
\address[F4E]{Fusion for Energy, Barcelona, Spain}

\begin{abstract}
  The ITER tokamak will be operating with 5 microwave diagnostic systems.
  While they rely on different physics, they share a common need:
  transmitting low and high power microwave in the range of
  \SIrange{12}{1000}{\giga\hertz} (different bandwidths for different
  diagnostics) between the plasma and a diagnostic area tens of meters
  away. The designs proposed for vacuum windows, in-vessel waveguides
  and antennas are presented together with the development
  activities needed to finalise this work.
\end{abstract}


\end{frontmatter}


\section{Introduction}
\label{sec:intro}

\paragraph{ITER environment} ITER is the first tokamak designed for
extensive operation using tritium and therefore producing high neutron
fluence. During the \emph{nuclear phase}, ITER is foreseen to produce up to
\SI{3e27}{neutrons} from up to \SI{15}{\kilogram} of tritium. This
nuclear environment has strong implications in the design of every
components of the tokamak. Radiological barriers ensures tritium
confinement in the main vacuum vessel. Neutron shielding protects
ex-vessel components from \SI{14}{\mega\electronvolt} neutrons and
limit their activation. In-vessel activation allows only maintenance
using remote handling techniques. All these constraints have to be
taken into account during the design of ITER microwave diagnostics and
their implications in term of the diagnostic performance assessed.

\paragraph{Reflectometers} ITER will have 3 different reflectometry
systems: Low Field Side Reflectometry (LFS-R), High Field Side
Reflectometry (HFS-R), Plasma Position Reflectometry (PPR)
\cite{vayakis-2006}. They are active systems emitting microwaves in
the range \SIrange{12}{170}{\giga\hertz} at low power (around
\SI{10}{\milli\watt}). They rely on the plasma property to reflect
microwave at a given position called cut-off layer. A probing
microwave is launched into the plasma and propagates up to the cut-off
layer where it is reflected. The wave travels back and is captured by
a detection system. From the phase-delay of the wave doing its
round-trip different physical properties can be extracted: plasma
electron density, position, turbulence and velocity
\cite{mazzucato-1998}. These diagnostics rely on an accurate
measurement of the complex reflected signal (phase and amplitude).

\paragraph{Electron-Cyclotron-Emission (ECE)} For temperature
measurement, an ECE system will be installed
\cite{udintsev-2011}. This is a passive diagnostic measuring the
microwave spectrum emitted by the plasma in the range
\SIrange{122}{1000}{\giga\hertz}. From the low frequencies,
the temperature profile is deduced. Higher frequencies (above 
\SI{350}{\giga\hertz}) are used for plasma radiation power
evaluation. A good signal amplitude is required to be able to
get a good accuracy.

\paragraph{Collective Thomson Scattering (CTS)} Fast ion velocity
distribution function will be measured by a CTS diagnostic. This
diagnostic relies on emitting a powerful (about \SI{1}{\mega\watt})
microwave beam at \SI{60}{\giga\hertz} into the plasma. The scattered
radiation is collected and recorded for different positions. This
system will emit a high power beam while measuring a very low power
radiation (in nW) \cite{bindslev-2007}.

All these diagnostics have in common to transmit microwaves from the
plasma to the outside world. The challenges reside in transmitting low
power signal within a hostile environment: tight space, important
heat fluxes, neutron fluence, large electromagnetic and seismic loads, stray microwave radiation and
ultra-high vacuum. While the microwave performance is a key aspect for
the design of the different diagnostic components, nuclear safety
remains the main concern.

This paper will address the design aspect and development activities
for the vacuum windows in section~\ref{sec:windows}, the in-vessel and
in-port waveguides in section~\ref{sec:WG} and the antennas
in~\ref{sec:antennas}. Finally conclusions will be drawn.

\section{Vacuum windows}
\label{sec:windows}
Microwaves are guided between the vacuum vessel and the
emission/detection equipment, located in a different building, using
oversized waveguides. To pass the vacuum barriers, vacuum-tight,
safety-important windows are used. On top of their role as vacuum
container, they also have to contain tritium in case of accidental
events (severe disruptions, earthquakes, fire\ldots). They are
critical in term of safety but also for the good performance of the
diagnostic: they have to be designed to minimise the microwave
reflection while keeping a robust and safe approach to meet the
requirements of a radiological barrier.

Nuclear regulation imposes some constraints on the window design. They
need to be able to sustain the usual loads as listed in
Section~\ref{sec:intro}. In term of pressure they have to be qualified
for a differential pressure of \SI{2}{\bar}. As a radiological barrier,
we have to assure they are not leaking. To do so, double windows are
installed with a monitored interspace. The interspace is pressurised
at about \SI{0.5}{\bar} with, potentially, a tracer gas. If the pressure
is observed to decrease, the leaking window is the plasma side one;
if the pressure increases, it signals an air-side leak. A tracer gas can
also be used in conjunction with a mass spectrometer to detect any
leak.

Two approaches have been followed to design such windows depending on
the space available to house them. When enough space is
available, the vacuum windows are placed on the vacuum boundary and
the microwave are transmitted across in a free space
propagation. When the space available is too tight to allow free
space propagation, the vacuum windows are placed inside the
waveguides, the later being the vacuum boundary.

\subsection{Free space propagation}
\label{sec:free-space}

\begin{figure}[t]
  \centering
  \includegraphics[width=0.2\textwidth]{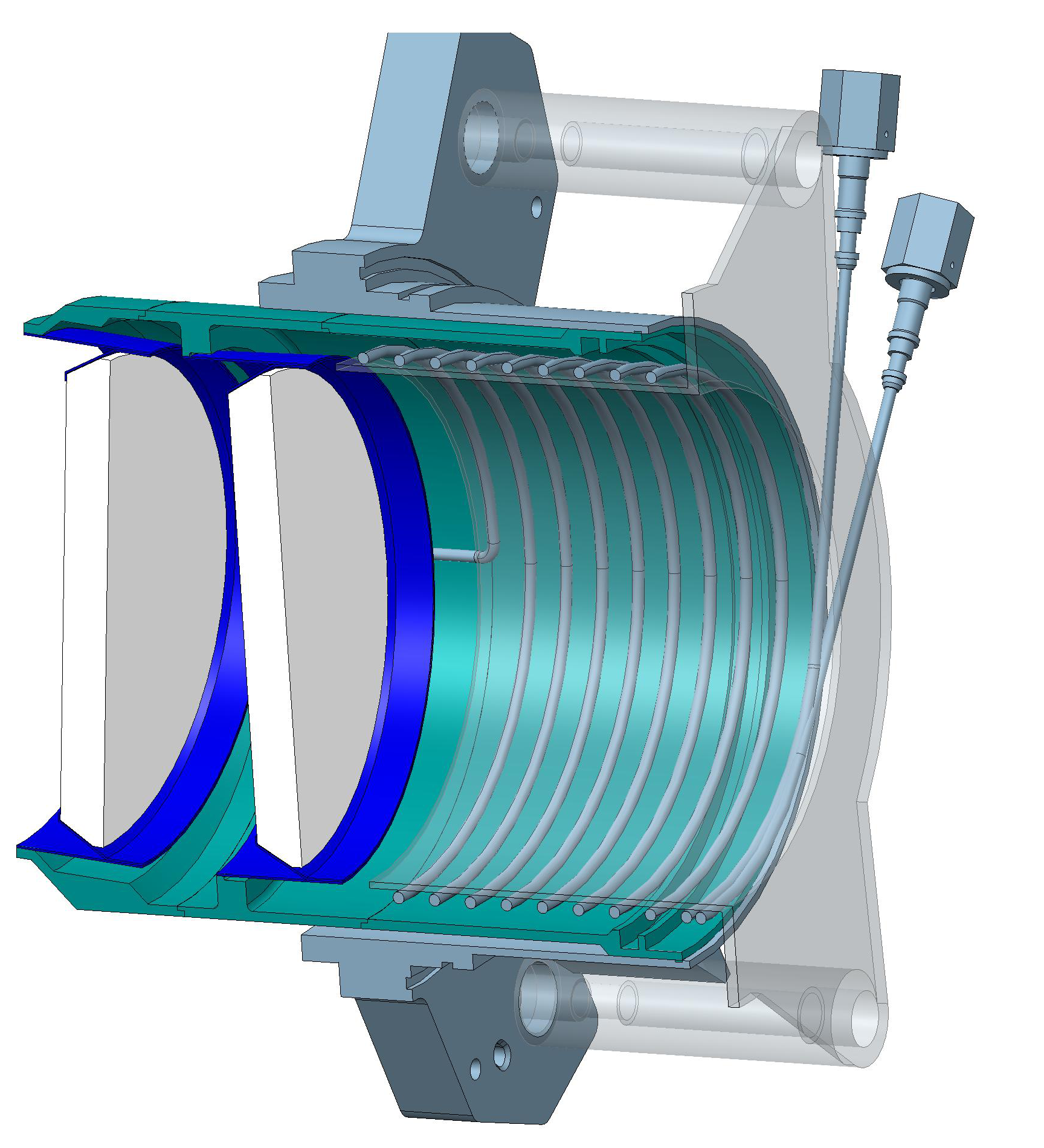}
  \caption{Double vacuum windows developed by ITER Organization for microwave diagnostics.}
  \label{fig:ece}
\end{figure}

This approach is inspired by what has been done on JET for the
reflectometry and ECE transmission lines~\cite{cupido-2005}. Bolted
and welded double window assemblies have been designed. They will be
installed on the closure plate at the back of the diagnostic
port-plugs as shown on Figure~\ref{fig:ece} for the ECE and LFS-R systems. A
Gaussian beam telescope will be used to transmit the microwaves
through the windows. Antenna on the air-side is aligned with the
vacuum-side optic (waveguide or mirrors) and the double vacuum windows
using elliptical mirrors. The impact of high power stray microwave radiation
\cite{oosterbeek-2014} on theses windows is being assessed.

\subsection{In-waveguide}
\label{sec:in-wg}

\begin{figure}[t]
  \centering
  \includegraphics[width=0.4\textwidth]{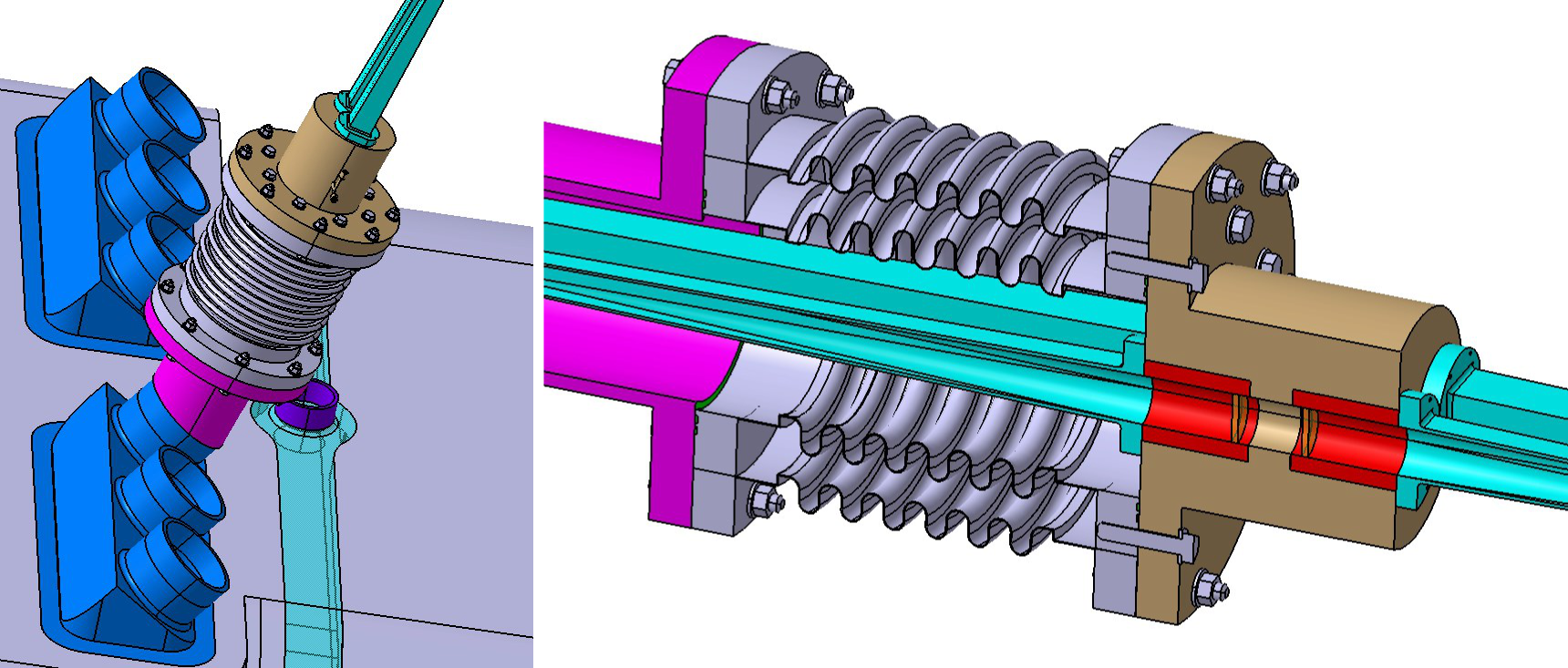}
  \caption{Concept for the reflectometry in-waveguide double
    windows developed by RF-DA. The dark blue part is the vacuum feedout and the double
    windows are placed in the red assembly.}
  \label{fig:inwg}
\end{figure}

In the case of in-waveguide vacuum barriers, the double
windows are embedded into the waveguide. Figure~\ref{fig:inwg} shows
one conceptual design to position the reflectometry in-waveguide
windows close to the upper-port feed-outs.

For this design, the microwave performance is checked by doing
numerical simulations. The losses simulated without any optimisation
were between 2 and \SI{5}{\dB}. Mock-ups are also being manufactured
for leak, thermal stress and microwave transmission tests.

\section{In-vessel waveguides}
\label{sec:WG}

\begin{figure}[t]
  \centering
  \includegraphics[width=0.4\textwidth]{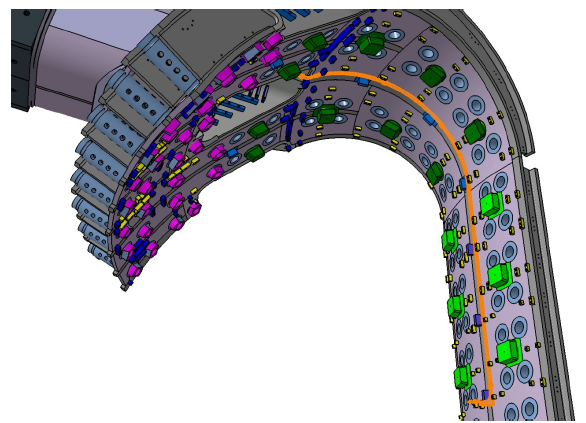}
  \caption{High Field Side reflectometry waveguides (highlighted) developed by RF-DA. The
    vacuum vessel is represented but the blanket modules are hidden
    for clarity.}
  \label{fig:wg}
\end{figure}

For HFS-R and PPR, waveguides are used to route microwaves between
windows and antennas through the vacuum vessel. Space available being
tight, the route has to be carefully designed in order not to
jeopardise measurements made through these small waveguides. Some
lines of sight are using an antenna on the inboard side of the vacuum
vessel. The waveguides are running for almost \SI{10}{\meter} along
the vacuum vessel from the upper ports of ITER as shown in
Figure~\ref{fig:wg}. The space available is limited as the nominal gap
between the vacuum vessel and the blankets is about
\SI{20}{\milli\meter}. Special cutouts have to be made on the back of
the blanket modules to allow the waveguides to run.

The internal section of the waveguides is smooth and rectangular, 12
by \SI{20}{\milli\meter} with a thickness of
\SI{1}{\milli\meter}. Traditionally such waveguides are in copper but
this cannot be done in ITER because of the large electro-magnetic
loads appearing during disruptions. To reduce the intensity of these
forces, stainless steel is used as its conductivity is much lower than
copper. In order to retain the microwave performance, a thin layer
(less than \SI{20}{\micro\meter}) of copper is deposed on the waveguide inner
surfaces. This design is a challenge to manufacture but some pieces
have already been produced by RF-DA. The copper adherence was good but microwave
performance is still being improved. The maximum length expected to be produced is about
\SI{2}{\meter}. Flanges will be used to assemble the complete
run.

Waveguides run in pairs as one is used for emission and one for
reception. A casing has also to be added in order to strengthen the
waveguides and allow them to cope the electro-magnetic loads. The
final section of the whole assembly is about 26 by \SI{80}{\milli\meter}.
To avoid mode conversion in order to keep
acceptable microwave performance, the waveguides must stay within the poloidal plane. 
Assembly will thus be challenging as the maximum deviation to this
plane is only \SI{0.3}{\milli\meter} per meter in the toroidal
direction.

\begin{figure}[t]
  \centering
  \includegraphics[width=0.3\textwidth]{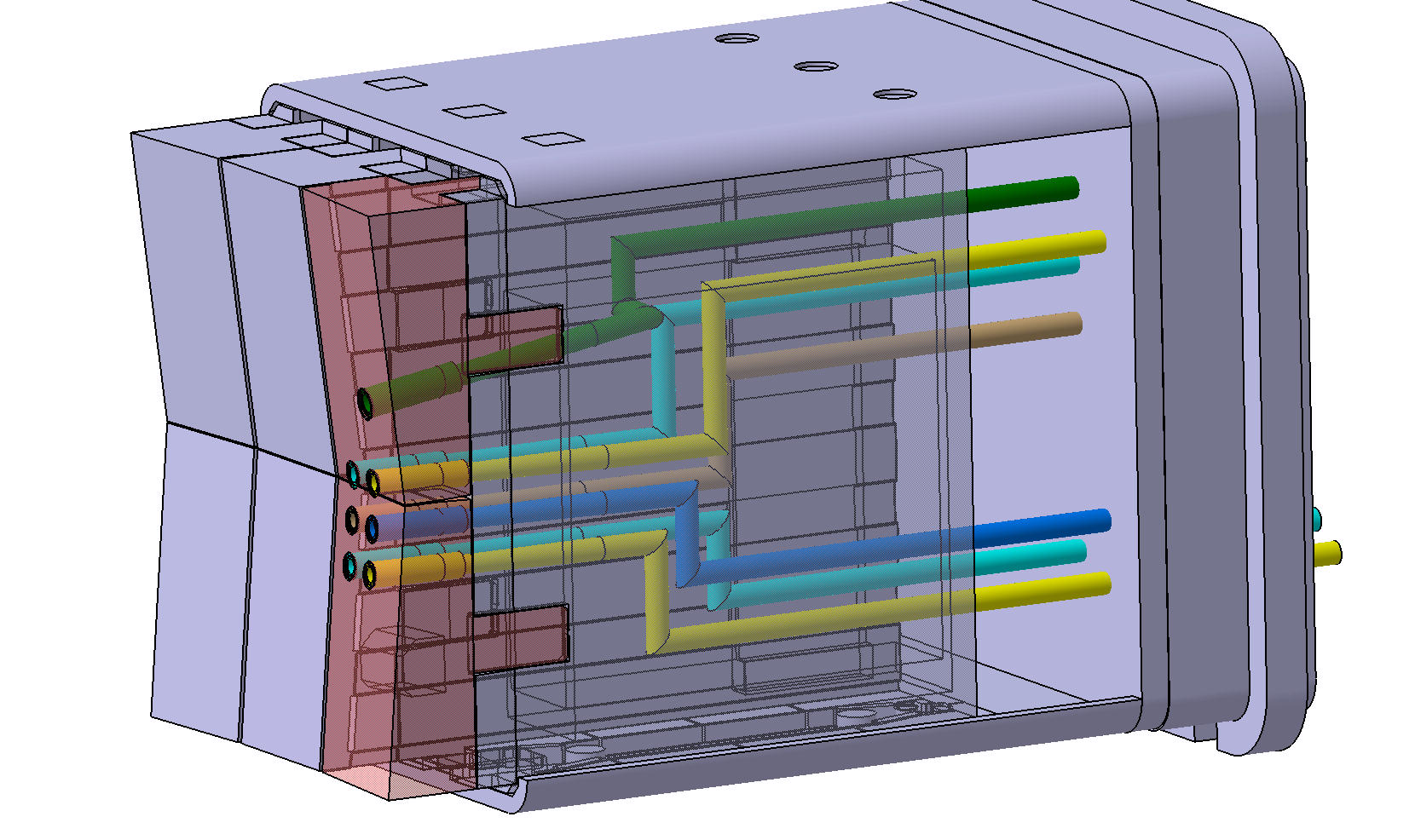}
  \caption{Low Field Side reflectometer antennas developed by US-DA. Oversized corrugated
    waveguide in the port plug are shown. The red part at the end of
    the port plug is the water cooled Diagnostic First Wall, handling the heat flux
    coming from the plasma.}
  \label{fig:lfsant}
\end{figure}

ECE, CTS and some reflectometry systems will also be installed in
diagnostic port-plugs \cite{pitcher-2012} where more space is
available. Depending on the system, quasi-optics or over-sized
corrugated waveguides will be used. The circular waveguide
inner-diameters will be 31.75, 50.8 or \SI{76.2}{\milli\meter}. Mitre
bends are used to create simple labyrinth in order to avoid straight
lines allowing neutron streaming as shown in
Figure~\ref{fig:lfsant}. Nevertheless, neutron fluence will induce
material activation and all the port plug components will be
maintained using remote handling tools.

\section{Antennas}
\label{sec:antennas}

\begin{figure}[t]
  \centering
  \includegraphics[width=0.3\textwidth]{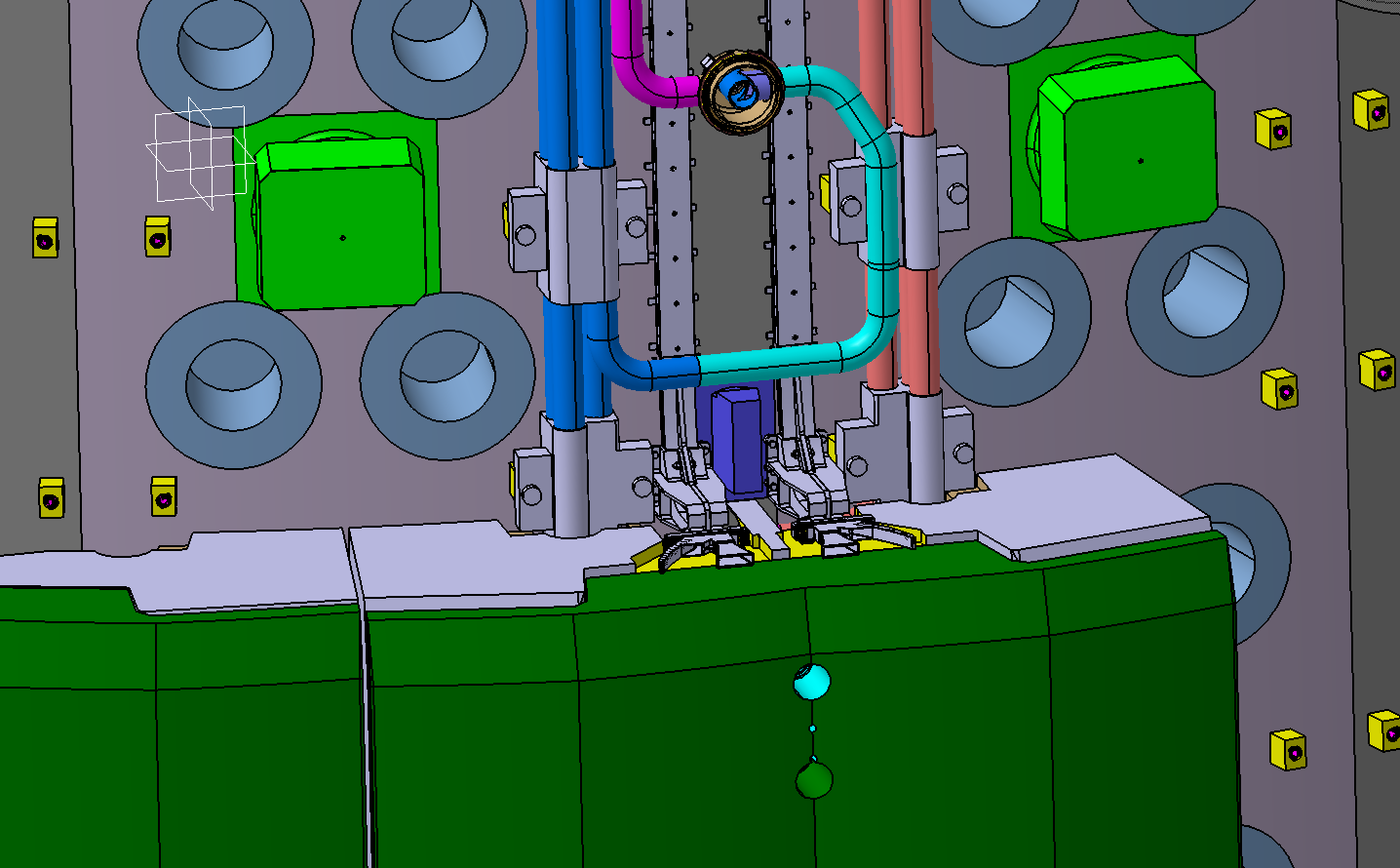}
  \caption{HFS-R antennas and waveguides. The top blankets have been
    hidden for clarity. When in place a \SI{40}{\milli\meter} clear
    gap is left for the diagnostic line of sight. The blue and red
    tubes are the cooling water pipes for the blanket systems.}
  \label{fig:hfsant}
\end{figure}

Depending on their position, antennas are facing different
challenges. For the inner board reflectometry systems, a
\SI{40}{\milli\meter} gap is allowed between two blanket modules in
order to get a clear view to the plasma. The design of the antenna and
its support has been constrained by the very tight space available as
seen in Figure~\ref{fig:hfsant}. The waveguide \SI{90}{\degree} bend
has been carefully designed to keep good microwave performance across
the whole frequency band (\SIrange{12}{140}{\giga\hertz}) and in both
polarisations. Some attention has also been taken to avoid having too
much current running in the bend structure during disruptions in order
to minimise the electro-magnetic loads.

Where more space is available, such as in the outboard side, collection optic can be
larger and recessed in the port plug. This is the case for the ECE and
CTS diagnostics. For reflectometry, more lines of sight are required,
and thus the last optics components cannot be recessed without posing
problem in terms of heat flux handling and neutron shielding. In this
case the waveguides are carrying the microwave up to first wall where
the antenna mouth are installed as shown in
Figure~\ref{fig:lfsant}. Being close to the plasma the antennas need
to be actively water cooled. Because of the amount of water required,
the design has to observe the French regulation on
pressurised equipment in a nuclear environment
(ESPN)~\cite{giacomin-2014}.

\section{Summary and future work}
\label{sec:conclusion}

While microwave diagnostic components require great attention for
their design in order to obtain good performance, ITER environment
brings additional constraints into the design: thermal and
electro-magnetic loads, neutron activation, tritium containment,
interface with other systems in a tight environment, microwave stray radiation and remote
handling maintenance. Antenna, waveguide and vacuum window designs
have followed these constraints. 

Most of the components described here are in an intermediate design
phase and are expected to go the final design phase in the next few year. Some
mock-ups have already been manufactured and tested to check the designs
fulfill the mechanical and thermal requirement, while keeping acceptable
microwave performance. This activity is expected to intensify in the
coming years. For the HFS-R and PPR systems, as they run along the
vacuum vessel, the main present activity is to consolidate the assembly
sequence in agreement with the blanket systems.

\textit{The views and opinions expressed herein do not necessarily
  reflect those of the ITER Organization. This publication reflects the views only of the author, and Fusion for Energy cannot be held responsible for any use which may be made of the information contained therein.}

\section*{References}

\bibliography{mybibfile}

\end{document}